          [2004/06/22 1.2 (KOF); 2001/04/25 1.1 (PWD)]

\documentclass[a4paper,twocolumn]{Gaia2004} 
\usepackage{times}      
\usepackage{epsfig}     
\usepackage{natbib}     
\title{Concluding Remarks: Gaia and Astrophysics in 2015--2020}

\author{Tim de Zeeuw}
\affil{Leiden Observatory}

\bibpunct{(}{)}{;}{a}{}{,}  

\begin{document}


\maketitle


\section{Introduction}

When Catherine Turon asked me to summarize this conference, I
responded positively because I thought it would be a good way to
understand all the recent progress on the Gaia project, which for the
past four years I had not followed very closely.  However, when I saw
the title assigned to me in the conference program, I was slightly
taken aback, as Catherine had neglected to mention that I was also
supposed to predict the future of astrophysics over the next ten to
fifteen years! Upon some further reflection, I realized that this is
of course precisely what is needed. It is crucial for any (space)
mission that takes of order a decade to construct, to check with some
regularity whether the original science goals are still considered
relevant, and whether the spacecraft specifications continue to make
it possible to achieve these goals.\looseness=-2

This was an exciting conference with many high-quality talks, and much
additional information in the form of posters, which were summarized
in the main session. All aspects of the development of the Gaia
mission were covered, from satellite status to data-analysis plans,
the activities of the fourteen working groups that support the Gaia
Science Team, and previews of the expected scientific harvest. The
award of a doctorate {\em honoris causa} to our `Nestor' Adriaan Blaauw
was a particularly noteworthy highlight. I will not attempt a detailed
summary, but will instead briefly review the key questions in
astrophysics, and the instrumentation on the ground and in space which
is being developed to answer these, and will then return to Gaia's
unique place in this overall scheme.\looseness=-2

\section{Predicting the Future}

The key astronomical questions of today are (i) the nature of dark
matter and dark energy, (ii) the formation and evolution of galaxies
from first light to the present, (iii) the physics of extreme
conditions (black holes and gamma ray bursts) and (iv) the formation
of stars and planets, and the origin of life.  The world-wide
astronomical community, together with national science foundations and
space agencies, aims to answer these questions by taking observations
with telescopes on the ground and in space, supported by
interpretative efforts and theoretical work.\looseness=-2

The major astronomical space observatories that are active this
current decade are the Hubble Space Telescope, the Spitzer Space
Telescope, the Chandra, XMM and Integral high-energy telescopes, and,
soon to come, the ESA Cornerstone Herschel/Planck. Missions dedicated
to specific topics include RXTE, SWIFT, ASTRO-F, COROT, Kepler, and
many others in the planning stage.

On the ground the 8--10~m class optical/infrared telescopes (Gemini,
Keck, Subaru, VLT) are being equipped with a full arsenal of
instruments, including many that take advantage of progress in
adaptive optics, and are also being linked interferometrically (e.g.,
VLTI) to obtain milli-arcsecond resolution. Numerous large-scale
surveys are available, including 2MASS, GSC-II, USNO-B, the 2dF and
Sloan Digital Sky Surveys, and more are being planned, including the
SDSS extension SEGUE to study Milky Way structure.  Many 2--4~m class
telescopes now focus on wide-field imaging surveys (MegaCam on the
CFHT, Omegacam on the VLT Survey Telescope, and VISTA). The RAVE
project aims to obtain millions of radial velocities for stars in the
Galaxy by multi-fiber spectroscopy. Following the success of MACHO,
EROS and OGLE, a new generation of synoptic facilities is being
planned, with PanSTARRS already under construction. LOFAR will provide
a major step forward for objects which emit extremely long-wavelength
radio waves.

The next decade will see the full power of the 8--10~m class
optical/infrared telescopes exploited with second generation
instruments and interferometric links, the completion of the first
ground-based global observatory ALMA, and the launch of SIM and the
James Webb Space Telescope. Construction of Extremely Large Telescopes
for the optical/infrared (GMT, TMT, and OWL), and the Square Kilometer
Array will start. The proposed 8m class Large Synoptic Survey
Telescope could provide deep imaging of (one hemisphere of) the entire
sky every four nights. It could well be complemented by imaging from
space to detect extrasolar planets by their microlensing signature
(MPF), and to detect supernovae for studies of dark energy (JDEM).
LISA will detect the gravitational wave signature of coalescing black
holes, and the very ambitious TPF--C mission could provide images of
extrasolar planets later in the decade. And finally, Gaia will provide
an all-sky photometric survey from launch in 2012, with the exquisite
astrometry and the radial velocities continuously improving until the
complete measurements will be available in 2018.\looseness=-2

\section{Gaia Science}

\subsection{Science case 2000}

The top-level summary of the Gaia science case as presented to ESA on
13~September 2000 in Paris stated the following. Gaia will determine:
{\parskip 0pt
\begin{itemize}
\itemsep -3pt
\parsep 0pt
\item when the stars in the Milky Way formed;
\item when and how the Milky Way was assembled;
\item how dark matter in the Milky Way is distributed.
\end{itemize}
Gaia will also make substantial contributions to:
\begin{itemize}
\itemsep -3pt
\parsep 0pt
\item stellar astrophysics
\item solar system studies
\item extrasolar planetary science
\item cosmology
\item fundamental physics
\end{itemize}
This is provided by a stereoscopic census of the Milky Way with the
following top-level specifications:\looseness=-2
\begin{itemize}
\itemsep -3pt
\parsep 0pt
\item astrometry for all objects with magnitude $G<20$ with accuracy 
      better than 10\,$\mu$as for the parallax, and 10\,$\mu$as/yr for the
      proper motion, for all stars brighter than $G=15$. This
      requires on-board detection, and is {\em the key} to the success
      of the mission (ASTRO);

\item radial velocities for (nearly) all objects with $G<17$, with an 
      accuracy between 2 and 10 km~s$^{-1}$ (depending on spectral
      type). These provide the crucial third component of the space
      motion of the objects, and are important for the measurement of
      perspective acceleration influencing the proper motions. The
      spectroscopy from which the radial velocities are derived also
      provides crucial astrophysical diagnostics of the objects (RVS);

\item broad- and medium-band photometry for all objects with $G<20$. 
      The broad-band measurements (BBP) are critical for chromatic
      correction, which otherwise would limit the astrometric
      accuracy. Together with the medium-band photometry (MBP) it
      provides astrophysical diagnostics, including an estimate of
      extinction, effective temperatures accurate to about 200 K, and
      [Fe/H] accurate to 0.2 dex.
\end{itemize}
A more detailed description of the Gaia science case and
specifications can be found in \cite{per01}.

}

\subsection{The Milky Way}

{\em Halo}. The key fossil record of the formation of the Milky Way is
to be found in the Galactic halo \citep{fbh02}. The overall aim is to
reconstruct the merger and accretion history from Gaia's unbiased
stereoscopic census by identifying the various star streams and
tagging them by the metal abundance and ages of the stars.  RAVE and
SDSS/SEGUE, if funded, will make good progress towards delineating
substructure in the Galactic halo. However, these surveys will provide
only one component of the three-dimensional space motion, and for 
$<1$\% of the stars to be measured by RVS.

{\em Disk}. The Gaia observations of stars in the Galactic disk will
provide an unprecedented record of the star formation history in the
disk of an average spiral galaxy. It will provide internal motions of
star forming regions and star clusters, the initial mass function, the
full census of binaries, and determine the age-metallicity relation in
the disk. The classical work by \cite{edv93} and \cite{nor04} provides
a small preview of what can be achieved in this way.

{\em Bulge}. The Galactic bulge is the nearest (barred?) spheroidal
galaxy component, and hence it is important to study in detail its
resolved stellar population and internal dynamics. Crowding and
extinction conspire to make only some windows available for study.
The Gaia measurements will cover a useful fraction of the Bulge, but
some of the science questions will be addressed earlier by studies in
very small fields, e.g., by combining multi-epoch HST imaging with
ground-based follow-up spectroscopy using multi-fiber or
integral-field spectrographs.

Gaia will provide estimates of ages and metallicities for many of the
stars it observes. These are crucial for unravelling the formation
history of the halo and the disk. The presentations at this conference
showed that much effort is still needed to try to improve the accuracy
of the ages and the [Fe/H]-values, and the effect of variations of
[$\alpha$/Fe] on these, as provided by the data from the MBP and the
RVS. Of course, once kinematic subgroups or streams have been
identified in the kinematics, their members can be followed up with
targeted higher-resolution spectroscopy from the ground. In all of
this the determination of the extinction is critical. This may well
need additional information from all-sky HI surveys and the 2MASS
infrared maps, and is likely to need further preparatory work with
wide-field imaging telescopes such as VST and VISTA.

{\em Dynamics}. A detailed model of the Galaxy needs to bring the star
counts and three-dimensional kinematics into harmony by using Newton's
laws of motion and gravity.  When applied to the full Gaia data this
approach has massive discriminating power, which will allow a
dynamical determination of the gravitational field of the Galaxy, and
provide the distribution of dark matter.\footnote{If it turns out that
Newton's law of gravity is modified at low accelerations
\citep{bek04}, the Gaia data will be equally useful.} James Binney
discussed three approaches to do this, and noted that while the
\cite{sch79} method is currently the most developed, with many
applications to model nearby galaxies, there are two alternatives
which merit further scrutiny, namely the `made-to-measure N-body
method' and the `Oxford torus machine'. Developing these methods to
the point where they can in fact deal with (do justice to) the entire
Gaia data set is a major undertaking, and needs to start now.  Much
experience in this area is available in Europe, and a coherent program
in this direction is an excellent example of a preparatory effort to
analyse and interpret the data that is beyond the scope of the Gaia
project itself, but is critical for the scientific success of the
mission. This is also an ideal project for further strengthening the
European Research Area, and lends itself naturally to network support
from the EU.

\subsection{Stars, the Solar system, Extrasolar Planets, Cosmology, 
            Transients, and Fundamental Physics}

{\em Stellar astrophysics}. Gaia's astrometric accuracy is such that
more than $10^7$ stars will have distances to better than 1\%, and
hence absolute luminosities to better than 2\% (provided the
extinction correction is accurate). As many as $10^8$ binary stars
will be detected, covering, at last, the full range of periods and
magnitude differences. Staffan S\"oderhjelm expects good orbital
solutions for $5\times 10^5$ systems, which will give accurate
masses. In combination with seismology from COROT, this will allow
absolute calibration of stellar models over the entire
Hertzsprung--Russell Diagram with unprecedented accuracy, as a
function of metallicity. This is a unique contribution of Gaia which
affects {\em all of} astrophysics, including the distance scale, but
more importantly, the calibration of stellar population models. The
VLTI will take the first steps in this direction, but will be limited
to at most a few hundred objects. As Yveline Lebreton made very clear,
a meaningful comparison with the observations will require much
preparatory work to further refine the (three-dimensional
hydrodynamic) stellar models, as many physical effects need to be
understood in detail.\looseness=-2

The Gaia photometric survey will reveal many known and new variable
stars, even though the cadence of observations is not optimized (nor
should it be) for the detection of such objects. Much of the
anticipated science may well be done before Gaia, by OGLE, PanSTARRS,
and the LSST. It will be difficult however for ground-based telescopes
to reach the exquisite homogeneity of the multi-band photometry that
Gaia will bring by virtue of its continuous scanning and its location
above the Earth's atmosphere. The preparation for the Gaia data on
variable stars will benefit from the investment in developments of
algorithms provided by, e.g., the OGLE collaboration.

{\em Solar system}. In the next decade much work on asteroids and
moons in the Solar system will be done from the ground.  Gaia's
astrometric accuracy will allow unique contributions to be made to the
dynamics of the Solar system, and will help identify Earth-crossing
dangers. The Gaia photometry of asteroids will surely be mined, but
work on variability will suffer from the limitations mentioned in the
previous paragraph.  The groups preparing for the analysis of the Gaia
data will also be in a good position to benefit from data of PanSTARRS
or the LSST.

{\em Extrasolar planets}. Gaia will provide a unique window on the
population of extrasolar planets in our Galaxy. Didier Queloz
demonstrated that the Gaia discovery space in the plane of semi-major
axis $a$ versus mass $m$ coincides with territory already explored
with the radial velocity method, transits, and, in the future, by
interferometry with VLTI/PRIMA and SIM. However, the complete Gaia
census is expected to provide accurate orbits for about 5000 systems,
more than an order of magnitude larger than can be expected by the
combination of all other methods in the same time frame. This makes
the Gaia contribution to this exciting and rapidly developing field
unique even in fifteen years time, and will allow a much increased
understanding of the population of planets as a function of the
properties of the parent star.

{\em Extragalactic science and cosmology}.  Gaia will measure fairly
accurate proper motions of stars in the nearby spheroidal satellites
of the Milky Way, and perhaps also in M31 and M33 by using globular
clusters and the brightest AGB stars. This will provide the dynamical
distance, internal orbit structure, and constrain the dark matter
content.  Ground-based work on nearby globular clusters has already
demonstrated the power of three-dimensional kinematics
\citep{ven05}. At cosmological distances, Gaia will provide a large
and homogeneous data set on quasars. These will define the global
reference frame and will also allow an independent determination of
the absolute three-dimensional acceleration of the Sun. The
astrometric selection of quasars might provide some surprises. It is
important to establish what other quasar science could be done beyond
the SDSS.

{\em Transients}. Wyn Evans discussed the possibility to detect
transient events with Gaia. This will require special `science alert'
software on the ground, which may require either {\em a priori\/}
information or use the Gaia data from the first half year of the
mission to establish the `steady' sky content to $G=20$ at Gaia's
spatial resolution. Given the limitations of the ground-station
coverage, with an anticipated data download once a day for about eight
hours, there will be limitations to the possibilities for rapid
follow-up of the fastest events. For transients discovered by other
means, it will be possible to go back to the Gaia data afterwards, and
check whether the on-board software in fact detected something. The
value of the expected detection of, e.g., as many as $5\times 10^4$
supernovae is hard to judge today. If JDEM is launched, it may provide
similar numbers well before 2018. Dedicated ground-based projects to
measure the dark energy parameters in other ways may also get there
before Gaia.

{\em Fundamental physics}. One of the aspects of the Gaia mission that
continues to impress me tremendously is the level to which the
astrometric measurements are influenced by general relativistic
effects at the $\mu$as level. This was beautifully illustrated by the
time sequence of the expected light-bending in the Solar system,
caused by the moving planets. Sergei Klioner's authoritative
presentation inspired much confidence that all these effects are
understood and can and will be incorporated in the data
reduction. This will allow measurement of various post-Newtonian
parameters with remarkable accuracy.

\subsection{Science case 2015--2020}

The top-level science goals for Gaia are unchanged from those
articulated in 2000. The study of the formation and evolution of the
Milky Way by means of a stereoscopic census with $\mu$as accuracy also
delivering stellar astrophysical parameters continues to be a crucial
complement to studies of galaxies at high redshift and in the nearby
Universe. Any science goals that require $\mu$as astrometry will
remain unique (e.g., the massive census of extrasolar planets). The
need for Gaia is arguably stronger than in 2000, because of the
unfortunate demise of DIVA and FAME, which would have done some of the
science, and the modest number of pre-selected targets to be observed
with SIM, albeit with great astrometric accuracy.

\section{The Gaia Observatory}

The core of the Gaia spacecraft is the ASTRO instrument.  The poster
with the real-size layout of the 1.5 Gpixel focal plane was an
impressive visual reminder of the technological challenge. The
presentations inspired confidence that the focal plane can be
assembled and tested, and will be able to deal with almost anything
that the sky offers (including the observation of the occasional
bright star without overloading the system). It is also clear that
minimizing radiation damage is critical.

The radial velocity spectrograph RVS is well-defined, and a simulator
is in place \citep{kat04}. Impressive progress has been made on the
ground with the RAVE precursor. This allows hands-on experience with
the Ca triplet spectral region, testing parts of the data analysis
software, including the recovery of stellar atmosphere parameters in
addition to the radial velocity, and already provides a preview of the
RVS science. This is an excellent way to prepare for the Gaia mission.

The details of the photometric system (BBP and MBP) are still under
discussion. It was interesting to hear of the work by the photometry
working group which now, finally, brings them close to a decision on
which bands to adopt. This should be done soon, so that the group can
then concentrate on optimizing the algorithms to get the highest
accuracy on the derived stellar parameters. It will be very useful to
equip a ground-based wide-field imager with the chosen filters, and
develop experience.

Much progress has been made in developing the complex software for the
on-board detection and data handling. Tests with OGLE and HST images
of crowded fields show very promising results, but the limited
on-board cpu power and the telemetry capacity remain a challenge. The
global iterative solution to obtain the astrometric measurements has
been shown to work. Much effort is devoted to steadily build up the
(simulation of) the end-to-end Gaia system, with many groups providing
components of the software. This entire effort is a textbook example
of a distributed European activity, with links to the development of
the GRID.

Finally, a few words about outreach. The Gaia project provides many
possibilities for education and outreach activities, from `The little
books of Gaia' to high-school exercises which attract students to the
physical sciences (see {\tt http://www.rssd.esa.int/Gaia}). The main
science goals are of general interest, in particular the search for
extrasolar planets.  The Gaia data also make it possible to simulate
traveling through the Galaxy, so that it is feasible to create the
legendary `astrometrics lab' from Startrek Enterprise, run by
Seven-of-Nine, 357 years ahead of its time (which relates to points
made by Jos de Bruijne and Xavier Luri during this conference).
Outreach is and should be an integral part of any (space) project.
Gaia in particular provides a key opportunity to go well beyond the
now standard press releases to in-depth efforts, supported by ESA
together with national programs, to engage the wider European public
in the excitement of scientific discovery.

\section{Gaia, a European Project}

This conference demonstrated that the initial Gaia concepts for
$\mu$as astrometry from space, developed in the mid-nineties by
Lennart Lindegren and Michael Perryman, and subsequently worked-out in
detail by the Gaia Science Advisory Group and approved by ESA, have
now led to a coherent effort by a well-organized, diverse, and
increasingly sophisticated team consisting of more than 250 European
astronomers.  The team combines the experience of senior astronomers,
some of whom helped make Hipparcos such a resounding scientific
success, with the energy and drive of young scientists and engineers.
The group works closely with ESA and the project has strong industrial
involvement.  The entire effort is overseen by the Gaia Science Team
under the eminent leadership of the project scientist Michael
Perryman.

The team has made much progress in the past four years. The three Gaia
instruments have been defined in great detail, and there is a much
improved understanding of the required algorithms and the full scope
of work. Much of this effort is funded locally. The Gaia science case
remains very strong and broad, but depends critically on maintaining
the specifications set in 2000, and summarized in Section 3. The
mission is very challenging---as an ESA Cornerstone should be---but
the team shows all the signs that it will be able to pull this
off. This will be a significant achievement for astronomy world-wide.

\section*{Acknowledgments}
It is a pleasure to thank Yves Viala, Catherine Turon, Michael
Perryman and Karen O'Flaherty for making this conference such a
tremendous success.

\end{document}